\documentclass[twocolumn,aps,prb,floatfix,showpacs]{revtex4}

\usepackage{graphics}

\begin{document}

%\widetext

\title{Evolution of Mid-gap States and Residual 3-Dimensionality in
La$_{2-x}$Sr$_x$CuO$_4$}

\author{S. Sahrakorpi$^{1}$, M. Lindroos$^{1,2}$,
R.S. Markiewicz$^{1}$ and A. Bansil$^1$}

\affiliation{
%\address{
$^1$Physics Department, Northeastern University, Boston,
           Massachusetts 02115, USA \\
$^2$Institute of Physics, Tampere University of Technology, P.O. Box
           692, 33101 Tampere, Finland}

\date{\today}

\begin{abstract}

   We have carried out extensive first principles doping-dependent
   computations of angle-resolved photoemission (ARPES) intensities in
   La$_{2-x}$Sr$_x$CuO$_4$ (LSCO) over a wide range of binding
   energies.  Intercell hopping and the associated 3-dimensionality,
   which is usually neglected in discussing cuprate physics, is shown
   to play a key role in shaping the ARPES spectra. Despite the
   obvious importance of strong coupling effects (e.g. the presence of
   a lower Hubbard band coexisting with mid-gap states in the doped
   insulator), we show that a number of salient features of the
   experimental ARPES spectra are captured to a surprisingly large
   extent when effects of $k_z$-dispersion are properly included in
   the analysis.

\end{abstract}

\maketitle

La$_{2-x}$Sr$_x$CuO$_4$ (LSCO) has drawn intense interest as a model
system for understanding one of the most hotly debated issues in
condensed matter physics currently, namely, how does a Mott insulator,
La$_2$CuO$_4$ (LCO), develop into a superconductor when doped with
holes via La/Sr substitution, and what is the route taken by the
electronic states in the insulator to achieve this remarkable
transformation into a metal.\cite{PR1,PR2} Angle-resolved
photoemission spectroscopy (ARPES) has been brought to bear on these
questions in recent years as techniques for preparing LSCO surfaces
have improved.\cite{zhou99,zhou01,yoshida01,ino02,yoshida03,zhou04}
ARPES spectra in LCO find a lower Hubbard band (LHB) associated with
an insulating state, which persists with finite hole doping, losing
intensity without shifting in energy.  At the same time {\em new
states} -- the so-called mid-gap states -- appear close to the Fermi
level, which evolve into the conventional metallic bands near optimal
doping.  Doping-dependent ARPES spectra in LSCO thus address a wide
range of issues concerning stripe and pseudogap physics and their
relation to the mechanism of high-temperature superconductivity, and
bear on questions of non-Fermi liquid behavior or gossamer
superconductivity\cite{laughlin02,zhang03}, among
others.\cite{damascelli03}

We have recently shown with the example of Bi2212 that the residual
$k_z$-dispersion of bands in a quasi-2D material will induce an
irreducible linewidth in ARPES peaks, which is unrelated to any
scattering mechanisms.\cite{bansilXX} This effect becomes accentuated
in LSCO, where the bands possess a greater 3D character compared to
Bi2212.  This article reports extensive first principles computations
of the ARPES intensity in LSCO with the goal of ascertaining the
extent to which $k_z$-dispersion affects the experimental spectra of
the mid-gap and LHB states.  The calculations properly model the
photoemission process and include the crystal wave functions to
describe the initial and final states in the presence of the surface
and take account of the associated dipole matrix element and its
dependencies on photon energy and polarization.  The interplay between
the effects of the ARPES matrix
element\cite{bansil99_2,lindroos02,sahrakorpi03} and the
$k_z$-dispersion yields computed intensity maps for emission from the
Fermi energy ($E_F$) as well as for binding energies several hundred
meV's below $E_F$, which are in surprising accord with the
corresponding measurements.

Our calculations give insight into a number of salient features of the
experimental spectra such as, the dispersion of the mid-gap states and
the characteristic broadenings and symmetries or lack thereof in the
photointensities. Evidence of physics beyond the framework of the
conventional local density approximation (LDA) based picture is
especially clear in the strongly underdoped regime, both in the
appearence of a d-wave-like pseudogap in the mid-gap states and in the
presence of the LHB near half filling.  We discuss this insulating
regime in terms of tight-binding (TB) computations, in which the
LDA-inspired overlap parameters are supplemented with a Hubbard $U$ or
a pseudogap $\Delta$.  The effect of $k_z$-dispersion is included
through an intercell hopping parameter -- to our knowledge for the
first time in connection with a TB description of the cuprates.  In
this way, via comparisons between theory and experiment, we adduce a
strong connection between the LDA generated metallic states and the
evolution of the mid-gap band throughout the doping range.

With regard to relevant computational details, the fully
self-consistent electronic structure of tetragonal LCO was obtained
within the LDA by using the well-established all electron Green
function methodology\cite{bansil99_1}; our band structure and Fermi
surface (FS) are in accord with published data\cite{yu87}.  Following
common practice, the metallic state of LSCO obtained by doping LCO
with Sr is assumed to be described by the LDA generated metallic band
structure for LCO. The effect of La/Sr substitution is mainly to
adjust the electron concentration and, therefore, we have modeled LSCO
at any given $x$-value by invoking a rigid band filling of the band
structure of LCO with the appropriate number of electrons per unit
cell.  All presented ARPES intensities have been computed within the
one-step photoemission formalism, assuming an LaO-layer terminated
surface; see Refs.~\onlinecite{bansil99_2,lindroos02}
and~\onlinecite{bansilXX} for details.

\begin{figure}
           \resizebox{6.5cm}{!}{\includegraphics{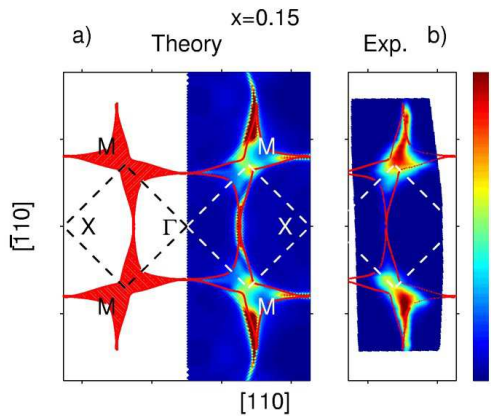}}
\caption{
    (Color) ARPES spectra for emission from the Fermi energy in
     optimally doped LSCO for $x=0.15$. (a): Red filled region on left
     hand side gives the 3D Fermi surface projected onto the
     $(k_x,k_y)$-plane, and denotes the area of allowed emissions.
     The right hand side gives computed intensity including the effect
     of the ARPES matrix elements with red lines marking boundaries of
     the red filled regions. (b): Corresponding experimental
     spectra\cite{zhou01}.
\label{fig:1}}
\end{figure}
Fig.~1 considers emission from $E_F$ for optimally doped LSCO and
shows the remarkable degree to which the FS map can be understood
within the framework of the conventional LDA-based picture. We start
by looking at the left side of~(a), where the filled red region gives
the projection of the FS onto the $(k_x,k_y)$-plane and encompasses
various FS cross-sections as a function of $k_z$.\cite{energywindow}
Emission of photoelectrons is possible in principle from any part of
this red region due to the quasi-2D nature of states. [In a strictly
2D system, the red region will collapse into a standard FS contour of
zero width.] We emphasize that the photointensity within this red
region will in general not be uniform as it will be modified by the
effect of the ARPES matrix element. This aspect is delineated on the
right side of~(a), where the theoretically predicted ARPES intensity
is shown superposed with the boundaries of the region of allowed
transitions by red lines.

When we compare the theoretical intensities in Fig.~1(a) with the
corresponding experimental results\cite{zhou01,zhouprivate} in
Fig.~1(b), the most striking feature is the appearance of wing-like
structures around the anti-nodal points $M(\pi,0)$ in both theory and
experiment. On the other hand, along the nodal direction in the first
Brillouin zone (BZ), $\Gamma$ to $X(\pi,\pi)$, the high computed
intensity is not reproduced in the measurements. However, we find that
this computed nodal intensity is quite sensitive to photon energy. It
will be necessary to carry out ARPES experiments over a range of
photon energies in order to ascertain whether or not absence of nodal
intensity represents a significant effect of correlations beyond the
LDA in LSCO.

\begin{figure}
           \resizebox{7.5cm}{!}{\includegraphics{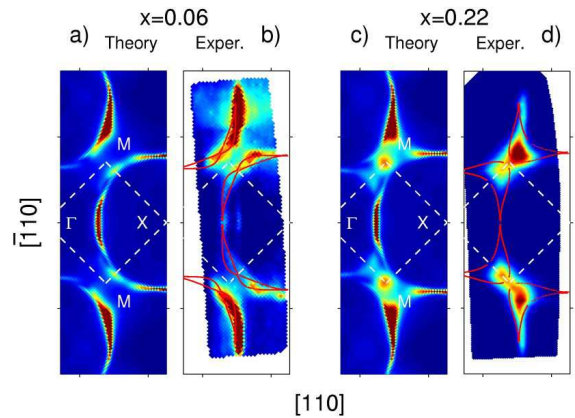}}
\caption{
    (Color) Similar to Fig.~1, except that this figure compares
    theoretical and experimental ARPES intensities in the underdoped
    $x=0.06$ [panels (a) and (b) (Ref.~\protect\onlinecite{zhou04})]
    and overdoped $x=0.22$ [panels (c) and (d)
    (Ref.~\protect\onlinecite{yoshida01})] cases.
\label{fig:2}}
\end{figure}
Figure~2 expands the preceding discussion to include the underdoped
and overdoped regimes. The computations in all cases simulate
experimental conditions\cite{zhou04,yoshida01} of resolution,
polarization of light and photon energy. The theory is seen to provide
a good overall description of the data over the entire doping range,
discrepancies along the nodal direction notwithstanding. Around the
antinodal point, incorporation of $k_z$-dispersion allows the
evolution of the FS across the (broadened) Van Hove singularity (VHS)
to be analyzed in detail.  In the underdoped sample the intensity is
concentrated in two features lying above and below the
$M(\pi,0)$-point. In contrast, the FS crosses the VHS near optimal
doping, and the overdoped sample displays a greater spectral weight
around the $M(\pi,0)$-point.  Since the midgap states evolve into
conventional bands with increased doping, the good agreement with
theory is perhaps to be expected in optimally and overdoped samples,
but the continued agreement for the underdoped sample is quite
remarkable -- particularly since the midgap states have a pseudogap,
as discussed below.

Figs.~1 and~2 make it clear that the broadening of ARPES spectra
resulting from the effect of $k_z$-dispersion is essentially zero
along the nodal direction and that it increases only gradually as one
moves towards the antinodal region. By contrast in the antinodal
region, the effects of the VHS conspire with those of the
$k_z$-dispersion to produce effective linewidths which increase
rapidly as one moves away from the antinodal point.  The antinodal
point itself is anomalous in that the broadening can be quite small
parallel to the BZ boundary ($X(\pi,\pi)$ to $M(\pi,0)$) as seen for
example from Fig.~2(a).  The present analysis shows that, even if
correlation effects were absent, features in the antinodal region
would generally be considerably broader than nodal ones, due simply to
the effect of $k_z$-dispersion.

Caution should be exercised in interpreting the significance of
differences between theory and experiment in Figs.~1 and~2 around say
the $M(\pi,0)$-point or for that matter even the nodal direction. On
the theoretical side, some of these details are sensitive to photon
energy and to the precise position of the Fermi energy in relation to
the VHS in the density of states. Similarly, higher resolution
experiments are needed to pin down some of the fine details in the
experimental spectra.\cite{asymmetry} Ultimately, ARPES evidence for
exotic physical effects beyond the conventional LDA-type pictures
(e.g. stripes, pseudogaps, strong correlations) must derive from a
careful analysis of the differences in the fine structure between
theoretical and experimental intensities such as those of Figs.~1
and~2.

\begin{figure}
           \resizebox{6.5cm}{!}{\includegraphics{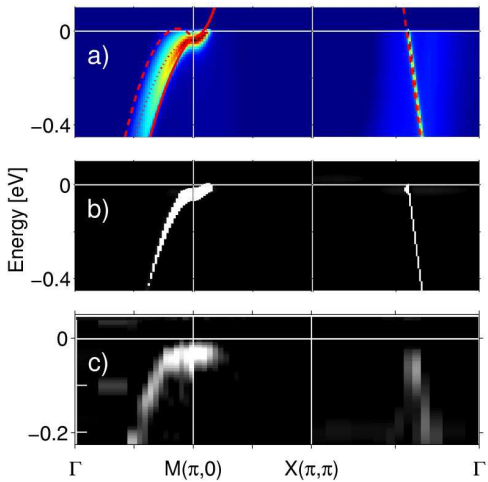}}
\caption{
    (Color) Band dispersion in optimally doped LSCO, $x=0.15$.~(a):
    Computed ARPES intensities along the $\Gamma-M-X-\Gamma$ -line in
    the BZ; color scale is the same as that in Fig.~1. Red lines give
    bands corresponding to different values of $k_z$: $k_z=0$ (solid
    line), $k_z=\pi/c$ (dotted), and $k_z=2\pi/c$ (dashed).~(b):
    Second derivative of the theoretical spectra in (a) as a grey
    scale plot.~(c): Experimental second derivatives\cite{ino02}
    corresponding to the theoretical results in~(b). Note scale change
    between panels~(b) and~(c).
\label{fig:3}}
\end{figure}
The typical behavior of spectra below $E_F$ is considered in Fig.~3
with the example of the optimally doped system. Much of the commentary
in connection with Figs.~1 and~2 above is applicable and need not be
repeated. In Fig.~3(a), the computed ARPES intensity along the
$\Gamma-M$ as well as the $\Gamma-X$-directions lies within the limits
given by the bands for various $k_z$-values, albeit with a strong
modulation by the ARPES matrix element.  Once again, the spectral
lines are substantially broader along $\Gamma-M$ compared to
$\Gamma-X$.  Theoretical and experimental\cite{ino02} results in~(b)
and~(c), respectively, are in substantial accord in this respect. Note
differences in vertical scale in~(b) and~(c), indicating a
renormalization (i.e. reduction) in bandwidth by roughly a factor of
two over the LDA predictions, which is quite common in the
cuprates. Also, the measured ARPES intensity in~(c) is seen to pull
away from the Fermi energy, which is not the case in the theoretical
plot of~(b). This difference however is to be expected since the
theory refers to the normal state, while the measurements are taken
from a superconducting sample.

We have fitted the first principles FS and band dispersions near $E_F$
in LSCO within the TB model using the form\cite{bansilXX}
\begin{eqnarray}
        \epsilon_k & = & -2t(c_x+c_y)-4t'c_xc_y -2t''(c_{2x}+c_{2y})
\nonumber \\
                   &   & \>\>\>
                      - 2 \> T_z(\vec k_{\parallel}) \> c_z{(c_x-c_y)}^2,
\label{eq:1}
\end{eqnarray}
with $c_i=cos(k_ia)$ and $c_{2i}=cos(2k_ia)$, $i=x,y$,
$c_z=cos(k_zc/2)$, and
\begin{equation}
         T_z = t_z \> cos{(k_x a/2)} \> cos{(k_y a/2)}. \label{eq:2}
\end{equation}
The extra angular dependence in Eq.~2 accounts for the staggered
stacking of neighboring CuO$_2$-planes, and results in vanishing
$k_z$-dispersion along the BZ boundaries.  The parameters which yield
a good fit are:
$t=0.32eV$, $t'=-0.25t$, $t''=-0.02t$, and $t_z=0.16t$.  The latter
value should be compared with $t_z=0.11t$ estimated from
transport.\cite{MK4} In order to simulate the presence of a pseudogap
and/or the superconducting gap, the bare bands of Eq.~1 should be
replaced by: $\epsilon_k \rightarrow \pm E_k$, where $E^2_k =
\epsilon^2_k + \Delta^2_k$ and $\Delta_k=\Delta_0(c_x-c_y)/2$. In this
way a $d$-wave-like gap with a maximum value of $\Delta_0$ at the
$M(\pi,0)$-point is produced in the spectrum; we find that $\Delta_0$
grows with underdoping as a pseudogap.

\begin{figure}
           \resizebox{6.5cm}{!}{\includegraphics{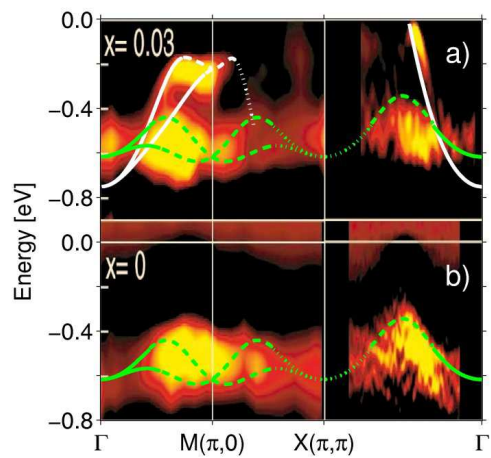}}
\caption{
    (Color) Theoretical band structures of LSCO are shown overlayed on
    experimental ARPES spectra (second derivatives) of
    Ref.~\protect\onlinecite{yoshida03} for two different dopings: (a)
    $x=0.03$, and (b) $x=0.0$. The data are given on a hot color scale
    in which red and black denote lows. Two sets of tight-binding
    bands are shown by green and white lines as detailed in the text.
    The two different bands in each case correspond to two values of
    $k_z$: $k_z=0$ and $k_z=2\pi/c$. Different line types denote
    computed relative spectral weights varying from 1.0-0.8 (solid
    lines), 0.8-0.2 (dashed) and 0.2-0.01 (dotted).
\label{fig:4}}
\end{figure}
Fig.~4 considers the underdoped regime below the
superconductor-insulator (SI) transition (around $x = 0.06$). Focusing
on the experimental\cite{yoshida03} spectra first, the half-filled
case ($x=0.0$) in~(b) is seen to display the presence of the lower
Hubbard band (LHB) at a binding energy of $\approx 0.5$ eV, and little
weight around $E_F$. When the system is doped with a small amount of
holes, we see from~(a) the appearance of a substantial weight in the
mid-gap spectrum near the $M(\pi,0)$ and $(\pi/2,\pi/2)$ points, while
the LHB is essentially unchanged.

The experimental results of Fig.~4 cannot of course be understood
within the conventional LDA-based picture, which fails to produce the
insulating state at half-filling. Insight can however be obtained via
the TB model, and accordingly, we have carried out two sets of such
calculations: (i) The LHB is fit to TB mean-field computations for a
saturated antiferromagnet with Hubbard parameter $U=6t$, with all
other parameters taken directly from the LDA bands. The
$k_z$-dispersion is accounted for via the $t_z$ parameter (see Eqs.~1
and~2), which to our knowledge has not been included in any previously
published work. These TB bands for the two extremal $k_z$-values are
given by green lines in Fig.~4, with different line types giving
associated spectral weights (see figure caption). Our parameter values
are consistent with earlier data from insulating cuprates based on the
ARPES\cite{KLBM} and spin wave spectra\cite{PeAr}. (ii) TB
calculations with the same parameters as those used in the preceding
case, except that the Hubbard parameter $U$ is replaced by a
$d$-wave-like pseudogap with $\Delta_0=220$ meV at $M(\pi,0)$, which
accounts for the presence of such a gap in the experimental spectra at
$x=0.03$.

The theoretical bands of Fig.~4 provide a handle on understanding some
salient features of the experimental spectra. We consider the LHB
along the $\Gamma-M$ and $M-X$ lines first. At both $x=0$ and
$x=0.03$, the measured LHB is seen to extend over the binding energy
range of 0.4-0.8 eV, more or less symmetrically about the
$M(\pi,0)$-point, at least insofar as its width is concerned. Much of
this width can be ascribed to the effect of $k_z$-dispersion, which
will induce spectral lines to spread between the boundaries given by
the pair of green bands. Note also that the computed LHB is symmetric
around $M(\pi,0)$ due to the effect of zone-folding in the AFM
insulating state.  The calculated spectral weights of the green bands
are seen to be larger along $\Gamma-M$ compared to the $M-X$
line.\cite{MEad} The experimental broadening at $M(\pi,0)$ is
anomalously large and not due to $k_z$-dispersion. This however is not
surprising since the $M(\pi,0)$-point lies on the AFM zone boundary
and is susceptible to `hot spot' scattering. The situation along the
$\Gamma-X$ line is sharply different. Here the effect of
$k_z$-dispersion is negligible. Moreover, $\Gamma-X$ is orthogonal to
the AFM zone boundary. Therefore, substantial observed widths of the
LHB along $\Gamma-X$ cannot be due to the effect of either
$k_z$-dispersion or of magnetic scattering.

We turn now to comment on the mid-gap spectrum in Fig.~4(a). Here the
aforementioned symmetry of the LHB with respect to the
$M(\pi,0)$-point is absent. The intensity of mid-gap states cuts off
quite abruptly near the $M$-point along the $M-X$ line. This is in
accord with the LDA-based bands (white lines), which show little
effect of $k_z$-dispersion and also possess little weight to the right
hand side of $M(\pi,0)$. On the other hand, a broad patch of intensity
is seen around $M(\pi,0)$ extending toward $\Gamma$.  This is to be
expected due to the effect of $k_z$-dispersion in view of the
boundaries given by the pair of white LDA-based bands.\cite{MEad}
Interestingly, along the $\Gamma-X$ -line, the experimental mid-gap
band is quite sharply defined, consistent with the fact that the LDA
bands display little $k_z$ induced broadening.

Taken together, the comparisons of Figs.~1-4 paint a remarkable
picture of the way mid-gap states evolve in LSCO with doping.  The
fact that the spectrum of the lightly doped insulator along the
$\Gamma-X$ line in Fig.~4(a) essentially follows the LDA bands,
suggests that even the first mid-gap states created when holes are
added into the Mott insulator mimic metallic states.\cite{Kyle}
However, the precise nature of the mid-gap states and how the
associated pseudogap evolves to yield a Luttinger-like Fermi surface
remains an important theoretical question.  The fact that hopping
appears only weakly renormalized with doping, but that the spectral
weight or the intensity of the mid-gap band scales with $x$, is much
in the spirit of a gossamer type model\cite{laughlin02,zhang03}.  The
presence of a pseudogap near $(\pi ,0)$ will result in predominantly
incoherent $c$-axis transport in the underdoped cuprates\cite{vdm},
whereas there may be coherent $c$-axis transport in overdoped
samples.\cite{hussey} Certainly, the ARPES spectra in the vicinity of
$E_F$ are described in considerable detail by the LDA computations as
seen from Figs.~1 and~2.

In conclusion, we have shown with the example of LSCO, that effects of
$k_z$-dispersion, which have been neglected in most of the existing
literature on the cuprates, play a key role in explaining the observed
broadening in the ARPES spectra throughout the BZ for both the mid-gap
band as well as the LHB. These results provide a new benchmark for
testing strong correlation models of the ARPES spectra. For example,
evidence for stripe or marginal Fermi liquid physics\cite{valla03}
must take into account the $\omega$-dependent broadening associated
with $k_z$-dispersion. Finally, despite the obvious importance of
strong correlation effects, the remarkable extent to which simple
LDA-type metallic states describe the dispersion of the mid-gap
spectrum in LSCO is an observation calling for theoretical
interpretation. Our results will provide a deeper understanding of
strong correlation effects in the cuprates, including insight into the
mid-gap states and the pseudogap, and will likely require some
revisions of associated models.

\begin{acknowledgments}

    This work is supported by the US Department of Energy contract
    DE-AC03-76SF00098, and benefited from the allocation of
    supercomputer time at NERSC, Northeastern University's Advanced
    Scientific Computation Center (ASCC), and the Institute of
    Advanced Computing (IAC), Tampere.

\end{acknowledgments}


\begin{thebibliography}{99}
\bibitem{energywindow}
%
         Following the experimental setup of integrating over narrow
         energy window, the projected FS's in Figs.~1 and 2 are
         integrated over relevant energy windows as follows: For
         $x=0.06$, $E_F \pm 5meV$; For $x=0.15,0.22$,
         $E_F-30meV+5meV$.
\bibitem{asymmetry}
%
         Note, for example, that the experimental spectra in
         Figs.~1(b),~2(b) and~2(d) are not completely symmetric with
         respect to the the upper and lower $M$-points.
\bibitem{MEad}
         There will of course be additional modulation of the spectra
         due to the ARPES matrix element, which has not been included
         in the TB computations of Fig.~4.
\end{thebibliography}
\end{document}